\documentclass[aps,prl,showpacs,twocolumn,amsmath,amssymb,superscriptaddress]{revtex4-1}
\usepackage[english]{babel}

\usepackage{graphicx,color}
\usepackage{amssymb}

\newcommand{\red}{\textcolor{black}}

\begin{document}

\title{\red{Hamiltonian ratchets with ultra-cold atoms}}

\author{Jiating Ni}
\affiliation{Department of Physics, Oklahoma State University, Stillwater, Oklahoma 74078-3072, USA}
\author{Siamak Dadras}
\affiliation{Department of Physics, Oklahoma State University, Stillwater, Oklahoma 74078-3072, USA}
\author{Wa Kun Lam}
\affiliation{Department of Physics, Oklahoma State University, Stillwater, Oklahoma 74078-3072, USA}
\author{Rajendra K. Shrestha}
\affiliation{Synrad Inc., 4600 Campus Pl., Mukilteo, Washington 98275, USA}
\author{Mark Sadgrove}
\affiliation{Research Institute of Electrical Communications, Tohoku University, Katahira 2-1-1, Aoba-ku, Sendai-shi Japan}
\author{Sandro Wimberger}
\affiliation{Dipartimento di Scienze Matematiche, Fisiche ed Informatiche, Universit\`{a} di Parma, Parco Area delle Scienze 7/A, 43124 Parma, Italy}
\affiliation{NFN, Sezione di Milano Bicocca, Gruppo Collegato di Parma, Italy}
\author{Gil S. Summy}
\affiliation{Department of Physics, Oklahoma State University, Stillwater, Oklahoma 74078-3072, USA}




\begin{abstract}
Quantum-resonance ratchets have been realized over the last ten years for the production of directed currents of atoms. These non-dissipative systems are based on the interaction of a Bose-Einstein condensate with an optical standing wave potential to produce a current of atoms in momentum space. In this paper we provide a review of the important features of these ratchets with a particular emphasis on their optimization using more complex initial states. We also examine their stability close to resonance conditions of the kicking. Finally we discuss the way in which these ratchets may pave the way for applications in quantum (random) walks and matter-wave interferometry.
\end{abstract}

\maketitle

\section{Introduction}
\label{sec:intro}
Ratchets are originally concepts of classical physics, relevant for the fundamental understanding of thermodynamics \cite{Feynman}, bio-molecular machines and microscopic pumps \cite{Motor1997, Reimann2002, AH2002, HM2009}. Quantum versions of ratchets have been proposed or implemented using cold atoms and Bose-Einstein condensates (BECs) 
\red{in optical potentials \cite{Tanja2002, Lundh2005, Gong2006, Gong2008, Renzoni2009, Pisa2006, Salger2009,DenisovPRA2007,GongPRE2004,Grossert_NatureComm2016}, or photons \cite{Zhang2015}}. These studies are based on a classical mechanism for producing a net current without net force, either by dissipation \cite{Luchinsky2000, Reimann2002, AH2002, HM2009} or in the case of so-called Hamiltonian ratchets by a proper choice of the initial conditions in phase space \red{\cite{DenisovPRA2007, Flach, Denisov2014, Ketz, GH2001}}. Hence the question arises what is really quantum about a {\it quantum} ratchet \cite{quantum}? The answer lies in the coherence properties of the wave-function, which not only determine the quantum motion, see e.g. \cite{Izr1990, wimberger2003quantum}, but also may be exploited for applications based on matter-wave interferometry. Here, we review previous implementations of atom-optics kicked rotor (AOKR) \cite{Raizen1999, sadgrovea2011experiment} ratchets \cite{Mark2007, Mark2007E, dana2008experimental, shrestha2012controlling, ni2016} and present a more comprehensive study on the stability, spread, and mean momentum of the ratchets. \red{Note that in this paper we do not examine systems in which a momentum current is generated in conjunction with a symmetry breaking of the ratchet potential \cite{Mark2013, fnl2013, Maarten2013, Gil2006}.}

We are investigating BECs kicked periodically in time at the Talbot resonance condition \cite{ryu2006high}, at the so-called quantum resonances of the quantum kicked rotor \cite{Izr1990, casati1984non, Raizen1999, wimberger2003quantum, wimberger2005experimental, sadgrovea2011experiment}. The quantum dynamics at those resonant conditions is known to be remarkably stable with respect to noise and perturbations, see e.g. \cite{fishman2002stable, wimberger2003quantum, wimberger2005experimental, sadgrovea2011experiment, Mikkel2009b, Gil2010, Maarten2014}. The stability of the ratchet is reviewed here for a broad range in the detuning of the drive from the exact resonance conditions, see also \cite{sadgrove2009pseudo, shrestha2012controlling, shrestha2013cold}.  All the ratchets described here are non-dissipative, and the symmetry breaking necessary for the production of a net current arises from choosing asymmetric initial states with respect to the kicking potential. This asymmetry can be enhanced by optimizing the initial state of the ratchets, as shown in \cite{ni2016} and elaborated below. Our experimental results highlight that in general a strong directed current and minimal spread go hand-in-hand with each other. This is important for potential applications that are sensitive to the quantum nature of the matter waves. Immediate applications are the realization of quantum walks in momentum space as recently proposed by two of us \cite{GW2015, GW2016} and interferometry with larger momentum differences as studied also in different contexts, see e.g. refs. \cite{mikkel2009a, mazzoni2015large, Kasevich}.

The paper is organized as follows: Section \ref{sec:2} reviews our earlier work \cite{ni2016} on AOKR ratchets created with superpositions of momentum eigenstates. We also present new experimental data showing how the mean momentum of the ratchet is affected by the number of these states. In addition, we show for the first time the influence of the relative phase between momentum states on the ratchet evolution. In Section \ref{sec:3} we report on the stability of the ratchet effect with respect to detuning from resonance in a so-called ``off-resonance" ratchet. Section \ref{sec:4} concludes the paper discussing possible future applications of optimized AOKR ratchets.

\section{Quantum-resonance ratchet}
\label{sec:2}

In this section, a theoretical analysis of the quantum ratchet is developed using a simple classical picture \cite{ni2016}. In addition, we discuss the experimental realization of the ratchets from initial atomic states prepared by Bragg diffraction of a BEC with pulsed optical standing waves. We show how one can control the ratchet momentum current with the phase and number of the initial momentum states.

\subsection{Theory}
\label{sec:2a}

To have a better understanding of the quantum ratchet dynamics, a review of the AOKR \cite{Raizen1999, sadgrovea2011experiment}, which is the building block of this phenomenon, is useful.
\red{The atom-optics realization of the paradigmatic kicked rotor model allows the experimental investigation of a fundamental, classically nonlinear system. On the quantum level, dynamical localization is perhaps the most celebrated phenomenon in the system \cite{FGP1982, Izr1990}. Dynamical localization occurs if the effective Planck's constant (which will be the dimensionless kick period $\tau$ below) is sufficiently irrational, and then the rotor's energy saturates in time. If, on the other hand, the effective Planck's constant is a rational multiple of $4\pi$ unbounded growth of energy typically occurs. This growth arises from resonant driving at the so-called quantum resonances \cite{Izr1990, casati1984non, sadgrovea2011experiment}, which do not have any counterpart in the corresponding classical system. In the following, we are focusing mainly on the dynamics at the principal quantum resonance $\tau=2\pi$ (this section) and its vicinity (section \ref{sec:3}).}

The theoretical framework of the quantum $\delta$-kicked rotor is constructed on the following dimensionless Hamiltonian \cite{graham1992dynamical}:
\begin{equation}
\label{ }
\hat{H}=\frac{\hat{p}^{2}}{2} + \phi_{d}V(\hat{x})\sum_{t=1}^{N}\delta(t^{\prime}-t\tau),
\end{equation}
where $\hat{p}$ is the scaled momentum in units of $\hbar G$ (two photon recoils), $\phi_{d}=\Omega^{2}\triangle t/8\delta_{L}$ is the strength of the kicks with $\triangle t$ the pulse length, $\Omega=\vec{\mu}\cdot\vec{E}(\vec{r})/\hbar$ the Rabi frequency, and $\delta_{L}$ the detuning of the kicking laser from the atomic transition. The experiments are conducted by exposing a BEC to $V(x) = \cos(Gx)$, which is the potential created by a pulsed optical standing wave where $G=2\pi/\lambda_{G}$ is grating vector. The spatial period of the standing wave $\lambda_{G}$ is determined by the direction of the laser beams and their wavelength. Here, $x$ is position, $t^{\prime}$ is the continuous time variable, and $t$ counts the number of kicks (pulses). $\tau=2\pi T / T_{1/2}$ is the scaled pulse period where $T$ is the pulse period in seconds and $T_{1/2}=2\pi M / \hbar G^{2}$ is the half-Talbot time for an atom with mass $M$. 
Due to the spatial periodicity of the kicking potential ($\lambda_{G}=2\pi/G$), only transitions between the momentum states that differ by integer multiples of two photon recoils, $\hbar G$, are allowed. Hence, $p$ can be broken down as $p=n+\beta$, where $n$ and $\beta$ are the integer and fractional parts of the momentum respectively. $\beta$ is a conserved quantity, and is also known as the quasi-momentum \cite{wimberger2003quantum, sadgrovea2011experiment}.

A picture of the ratchet mechanism can be developed from a consideration of the gradient of the standing wave, which serves as a driving force on the wave function of the atoms. With this in mind, one can deduce that maximizing the wave function intensity at positions with larger potential gradients should result in a greater net force and accordingly higher possibility of ratchet creation. In this picture, the sign of the potential gradient near the wave function's maxima will determine the direction of the ratchet. In order to create the spatially localized atomic wave function, an initial state composed of two or more plane waves can be utilized:
\begin{equation}
\label{ }
|\psi\rangle=\sum_{n}e^{-in\gamma}|n\rangle,
\end{equation}
where $|n\rangle$ signifies the state $|n\hbar G\rangle$, and $\gamma$ is an offset phase. As will be seen shortly, when $\gamma=\pi/2$ the maxima of the spatial wave function are shifted to where the potential gradient is greatest. It is also worth mentioning that setting the offset phase of the wave function is identical to setting the offset phase of the potential to be $\gamma$.

Assuming that a single BEC has a narrow momentum width, its wave function in momentum space can be represented as $\psi(p)=\delta(p)$, where $p$ is the continuous momentum variable. This wave function can be Fourier transformed into position space to be studied in the frame of the standing wave. In this case the wave function in position space is a plane wave with a uniform spatial distribution ($|\phi(x)|=\sqrt{G/2\pi}$). According to the simple picture described above, no ratchet will be formed due to the absence of a net force on the atoms. 

In contrast, if the initial state contains more than one plane wave, the possibility exists for the creation of a ratchet. The wave function of such a state can be written in position space as:
\begin{equation}
\label{ }
\phi(x)=A\sum_{n}e^{in\gamma}e^{ip_{n}x/\hbar},
\end{equation}
where $A$ is a normalization factor. In Fig. 1 we plot the population spatial distribution, $\left|\phi(x)\right|^{2}$ with $\gamma=\pi/2$, and the standing wave potential. Fig. 1(a) illustrates the population distribution for superpositions of two and seven consecutive plane waves. Note that peaks of the population in Fig. 1(a) arise at positions where the standing wave happens to have the greatest gradient. As expected from a Fourier transform, when the initial state is composed of more plane waves it becomes more spatially localized. That is the dashed line, corresponding to seven plane waves, shows a higher maximum intensity and a noticeably reduced full width at half maximum (FWHM). This can also be inferred from Fig. 2(a), which shows directly that by increasing the number of consecutive plane waves in the initial state (that is, increasing the range of the momentum states of an initial state), the population peaks' FWHM decreases. These narrower wave functions not only experience a larger average value of the potential gradient, but also less gradient variation. Therefore a ``cleaner'' ratchet would be expected at higher number of plane waves in the initial superposition. \red{We also expect that to optimize the ratchet current, the initial wave function should contain a superposition of consecutive momentum states. For example, the superposition $|-1\rangle+|1\rangle$ would be a less optimal ratchet state than $|0\rangle+|1\rangle$.} Fig. 1(b) illustrates the spatial distribution of the atomic population for initial states with different superpositions of nonconsecutive momentum states. Note that although some population peaks appear at greatest gradient positions, there are a considerable number of peaks that arise at positions where the gradient is lower, zero or even of opposite sign. This helps explain why the ratchet is weak or even absent in experiments starting with nonconsecutive initial states. Another interesting feature of Fig. 1(b) is the behavior of the initial state $e^{-i\pi/2}|-1\rangle+e^{i\pi/2}|1\rangle$ (dash-dotted line). This initial state gives population peaks at positions where the positive and negative gradient is greatest suggesting that one should expect two ratchets at opposite directions.
\begin{figure}
\begin{center}
\includegraphics[width=0.5\textwidth]{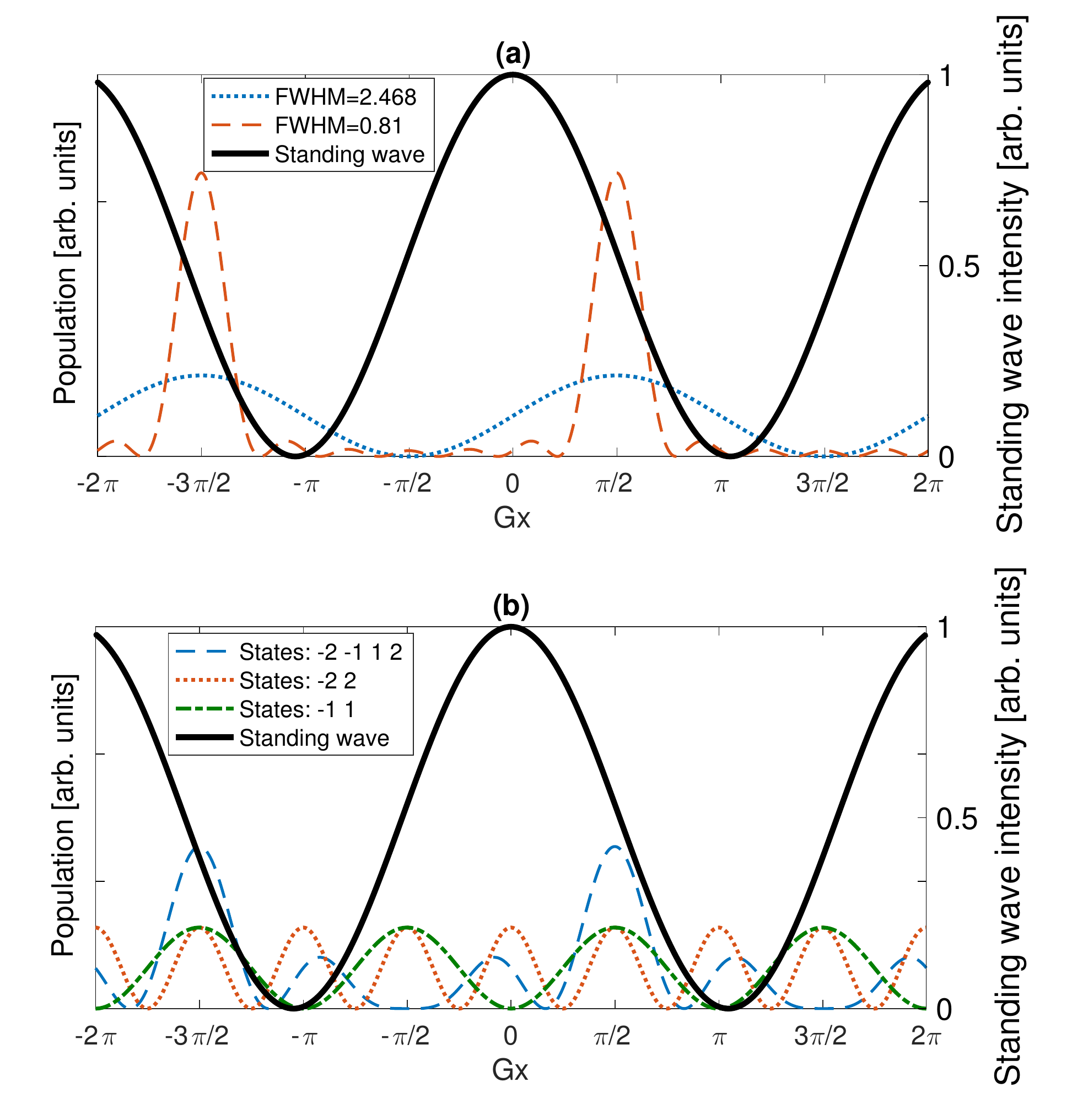}
\caption{Panel (a) shows the atomic population for superpositions of two, $|0\rangle+ e^{i\pi/2}|1\rangle$ (dotted line), and seven, $\sum_{n=-3}^{3}e^{-in\pi/2}|n\rangle$ (dashed line) plane waves. Panel (b) shows atomic population for initial states composed of superpositions of nonconsecutive momentum states: $e^{-i\pi}|-2\rangle+e^{-i\pi/2}|-1\rangle+e^{i\pi/2}|1\rangle+e^{i\pi}|2\rangle$ (dashed line), $e^{-i\pi}|-2\rangle+e^{i\pi}|2\rangle$ (dotted line), and $e^{-i\pi/2}|-1\rangle+e^{i\pi/2}|1\rangle$ (dash-dotted line). For both panels, the solid lines represent the spatial distribution of the standing wave intensity. \red{From \cite{ni2016}}.}
\label{ }
\end{center}
\end{figure}

To help quantify these ideas we define the effective force on the atoms due to the standing wave as
\begin{equation}
\label{ }
F_{\textrm{eff}}=\left|\int_{-\pi}^{\pi}\left|\phi(x)\right|^{2} \frac{dV(x)}{dx}dx\right|.
\end{equation}
This function is shown for different initial states in Fig. 2(b). The solid line reveals an enhancement of $F_{\textrm{eff}}$ increasing with the number of consecutive plane waves in the initial state. The dashed line is the effective force for the same momentum state range with one state missing. Note that the missing momentum state can be any except the first and last momentum states of a initial state. To ensure the peak of the atomic spatial distribution appears at a position with maximum $F_{\textrm{eff}}$ (in this case $Gx=\pi/2$), choosing the proper phase for each plane wave is crucial. Thus the phases should be set to $e^{-in\pi/2}$, where $n$ is the momentum state index. If the phases differ from these values the peak of the atomic spatial distribution will shift away from the greatest gradient and the effective force will become weaker. Fig. 2(c) illustrates the effective force of the standing wave as the function of the offset phase $\gamma$. Effective force is maximized when $\gamma=\pi/2$, which perfectly matches Fig. 1(b). The other peak of $\gamma=-\pi/2$ represents the same ratchet current but in the opposite direction.
\begin{figure}
\begin{center}
\includegraphics[width=0.5\textwidth]{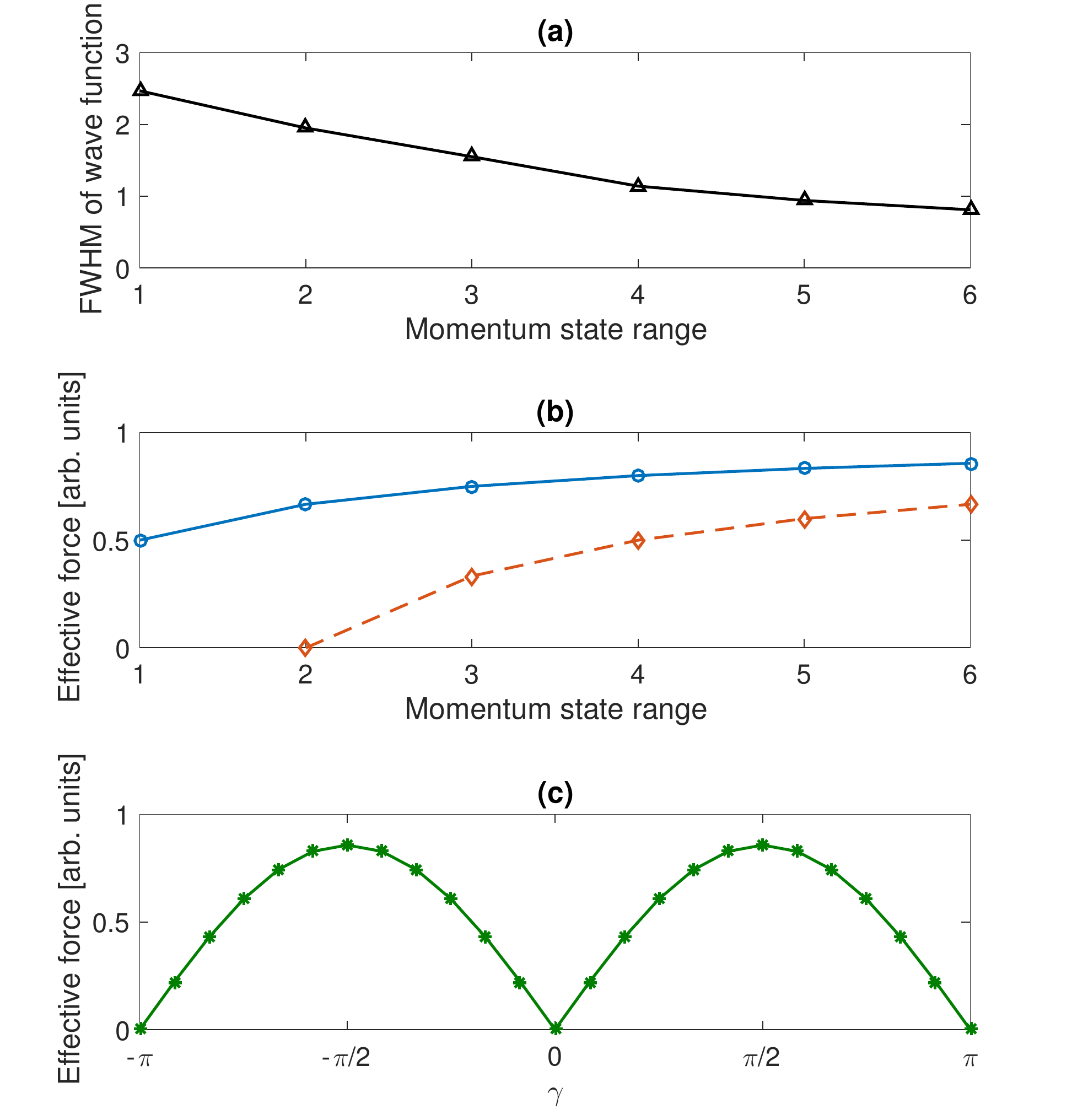}
\caption{Panel (a) shows theoretical data for the atomic spatial distribution FWHM as a function of the number of consecutive plane waves in the atomic state. \red{Panel (b) shows the effective force of the standing wave as a function of the momentum state range. The circles represent the case where the momentum states are consecutive, diamonds the situation where one state is missing within the range.} Panel (c) shows the effective force of the standing wave as the function of the offset phase $\gamma$ with a initial state containing seven consecutive momentum states $\sum_{n=-3}^{3}e^{-in\pi/2}|n\rangle$. }
\label{ }
\end{center}
\end{figure}

 \begin{figure}
 	\begin{center}
 		\includegraphics[width=0.5\textwidth]{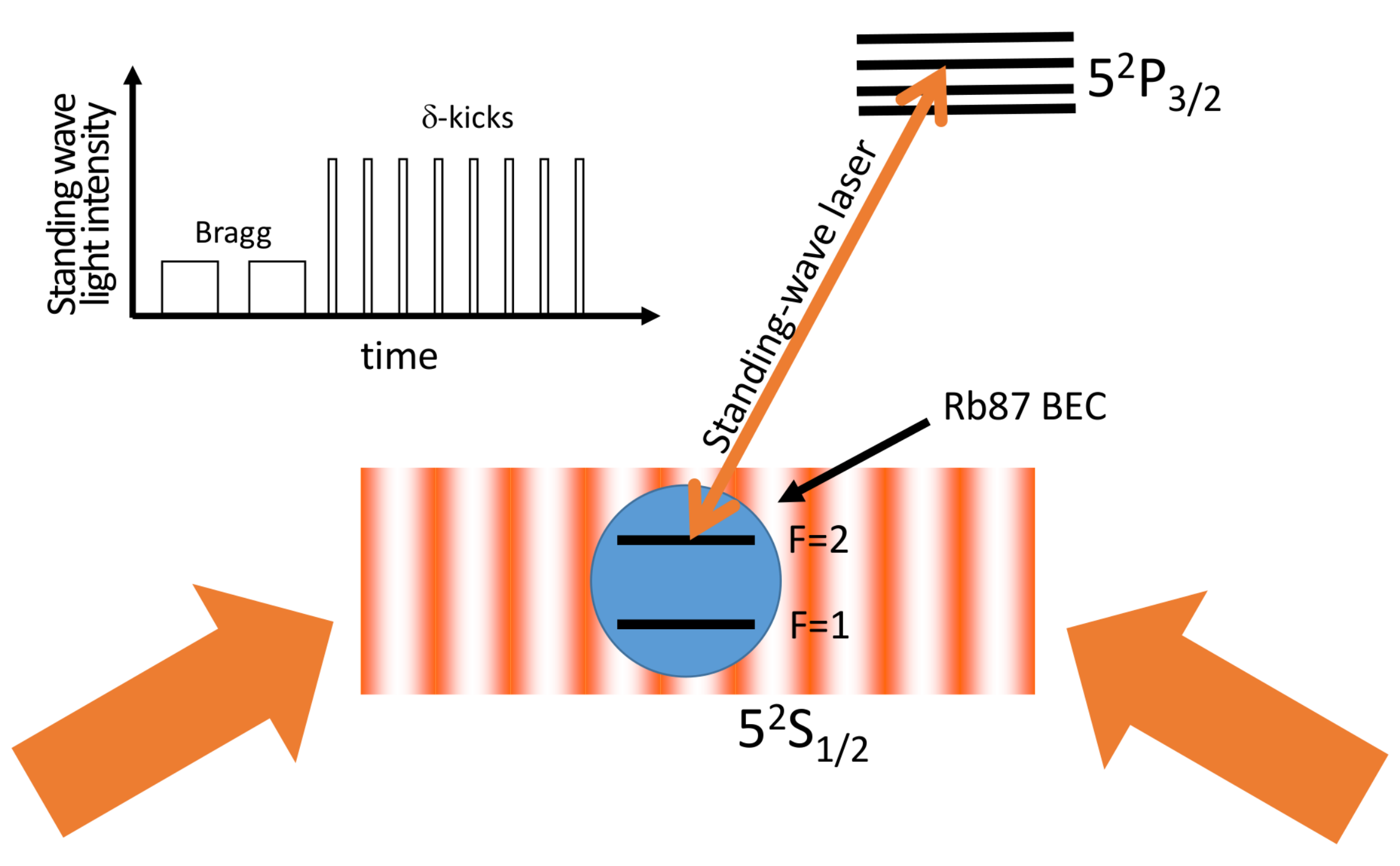}
 		\caption{\red{Schematic showing the geometry of the experiment and the sequence of Bragg and kick pulses used to study quantum ratchets.}}
 		\label{ }
 	\end{center}
 \end{figure}

\subsection{Experimental apparatus}
\label{sec:2b}

To conduct the ratchet experiments, a BEC of about 70,000 $^{87}$Rb atoms in the $5S_{1/2}$, $F=1$ hyperfine ground states was created by evaporative cooling of optically trapped atoms, held in a focused CO$_{2}$ laser beam. This procedure, including details of our MOT, optical molasses cooling and evaporative cooling are described elsewhere \cite{barrett2001all}. The ratio of the mean field energy to the recoil energy in our experiment is approximatively $10^{-5}$, which puts the dynamics in a region where the nonlinearity is negligible. Immediately after the BEC was released from the trap, a series of horizontal optical standing wave pulses were applied, providing the Bragg and regular $\delta$-kicks in the sequence needed for each experiment (see Fig. 3). The standing wave was created by two laser beams with a wavelength of $\lambda=780$ nm, which was 6.8 GHz red detuned from the $5S_{1/2}$, $F=1$ $\longrightarrow$ $5P_{3/2}$, $F'=3$ transition. Each laser beam was aligned $53^{\circ}$ from the vertical giving a standing wave spacial period of $\lambda_{G}=\lambda/(2\sin53^{\circ})$. This led to a half-Talbot time of  $T_{1/2}\approx 51.5$ $\mu$s, with the primary quantum resonances falling at integer multiples of this time \cite{ryu2006high}. Standing wave pulses were created by controlling the phase, intensity, pulse length, and the relative frequency between the two laser beams. This was realized by running the standing wave's constituent laser beams through acousto-optic modulators (AOMs), each of which driven by an arbitrary waveform generator. The nodes of the standing wave were displaced with velocity $v=2\pi\triangle f/G$ with $\triangle f$ being the frequency difference between the two beams. Therefore the momentum of the BEC, $p$, relative to the standing wave could be controlled through $\triangle f$ due to its proportionality to the standing wave velocity. The duration of the kicking pulses was fixed at 600 $n$s to ensure that the experiments were performed in the Raman-Nath regime \cite{gould1986diffraction}. That is, the evolution of the wave function due to its kinetic energy was negligible during the pulse.

To prepare an initial state composed of a desired superposition of several plane waves, a sequence of longer standing wave pulses were applied in the Bragg configuration \cite{nath1936diffraction,kozuma1999coherent}. Each Bragg pulse couples two momentum states with an interaction matrix given by
\begin{equation}
\label{ }
U=\left(\begin{array}{cc}
  \cos(\frac{\Omega_{B}\tau_{B}}{2})    &  -i\sin(\frac{\Omega_{B}\tau_{B}}{2})\exp(i\gamma_{B}) \\[3mm]
  -i\sin(\frac{\Omega_{B}\tau_{B}}{2})\exp(-i\gamma_{B})    &   \cos(\frac{\Omega_{B}\tau_{B}}{2})
\end{array}\right),
\end{equation}
in which $\Omega_{B}$ is the effective Rabi frequency, $\tau_{B}$ is the pulse length, and $\gamma_{B}$ is the offset phase of the standing wave used for a Bragg diffraction. The procedure of preparing a desired superposition of momentum states using Bragg pulses can be understood by examing the example for the preparation of five consecutive states shown in Fig. 4; the first Bragg pulse with $p=0.5$ diffracts atoms from momentum state $|0\rangle$ to $|1\rangle$. Then the second pulse with $p=-0.5$ diffracts atoms from $|0\rangle$ to $|-1\rangle$, leaving the atoms at state $|1\rangle$ untouched. Similarly, the third and fourth Bragg pulses, with $p=1.5$ and $p=-1.5$, diffract atoms from $|1\rangle$ and $|-1\rangle$ states to $|2\rangle$ and $|-2\rangle$ respectively. There is considerable freedom in the Bragg pulse duration. However in our experiment this parameter was set to 103 $\mu$s (one Talbot time) to ensure that the phases of all prepared states were preserved during the time in which the subsequent Bragg pulses were applied. In addition, the intensity of each Bragg pulse was adjusted to obtain equal populations in all states. The relative phases between the Bragg-prepared states are essential to achieve the expected dynamics of the ratchet. Therefore, the experiments were conducted so that all of the phases of the states in the superposition were identical. This was achieved by setting $\gamma_{B}=-\pi/2$ or $\gamma_{B}=\pi/2$ depending on whether states with $\Delta n=-1$ or $\Delta n =1$ were being coupled.
\begin{figure}
\begin{center}
\includegraphics[width=0.5\textwidth]{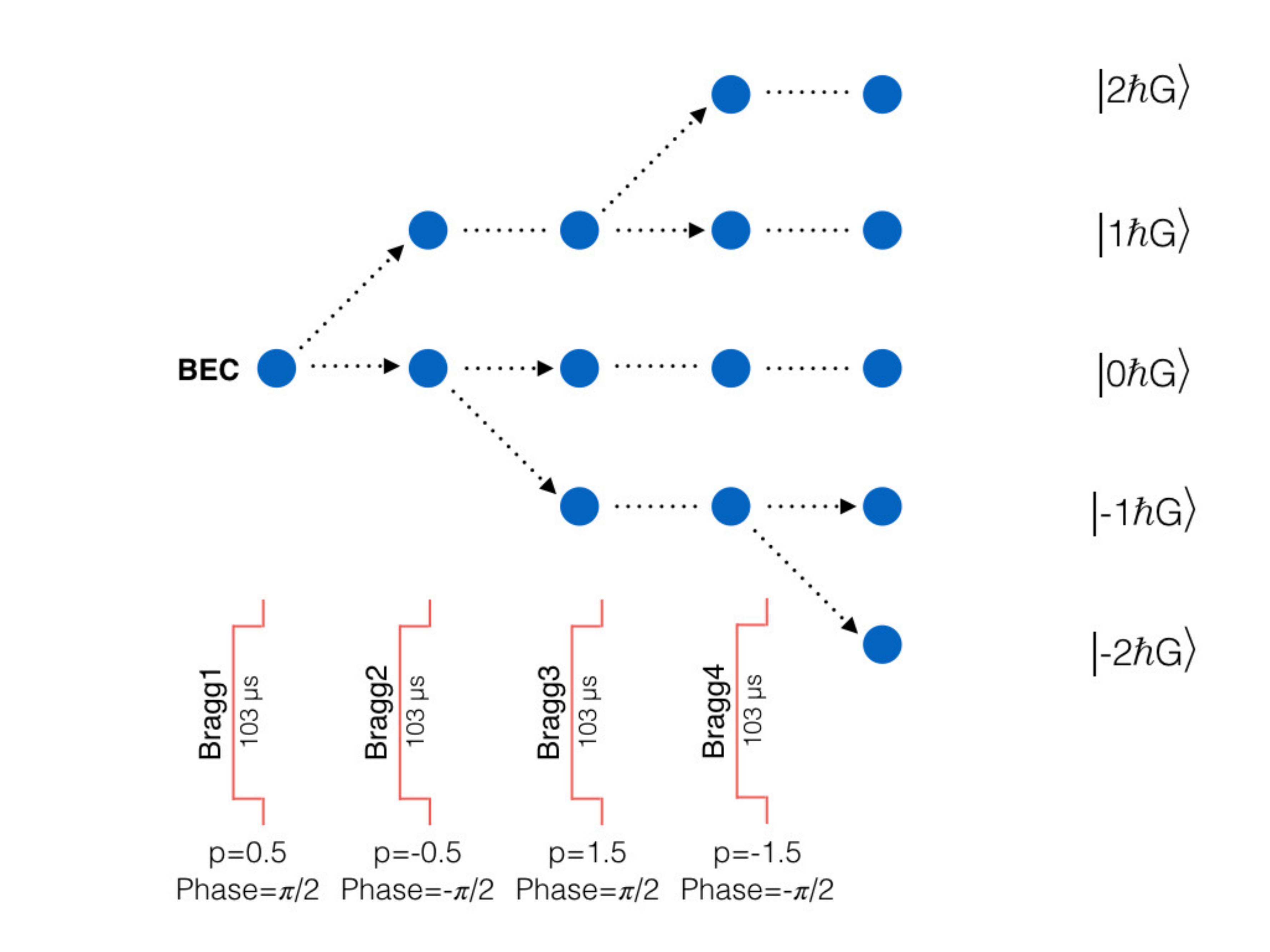}
\caption{Experimental schematic for preparation of the five component initial state $\sum_{n=-2}^{2}|n\rangle$ using four Bragg pulses. From \cite{ni2016}.}
\label{ }
\end{center}
\end{figure}

Following the Bragg preparation of the initial state, the $\delta$-kick rotor pulses were immediately applied. These short pulses diffracted the atoms into a wide range of momentum states. For the kicking pulses, the pulse strength $\phi_{d}$ $\sim$ 1.4, and the phase $\gamma$ in the potential was $\pi/2$. The latter is mathematically equivalent to individually setting the phases of each initial momentum states to be $e^{in\pi/2}$ which is required to optimize the ratchet current (see Fig. 2(c)). For the resonant ratchet experiments, the time between kicking pulses was 51.5 $\mu$s (half-Talbot time) with initial momentum $\beta=0.5$ to maintain quantum resonance. In order to measure the momentum distribution, after the Bragg and $\delta$-kick pulse trains were applied, atoms were absorption imaged following a free flight time of 10 ms. Examples of time-of-flight images for three different ratchets are shown in Fig. 5. Of particular note is Fig. 5(c) for the initial state $|-1\rangle + |1\rangle$ where two oppositely directed ratchets can be seen. This can be understood using the idea of Fig. 1(b) where two identical peaks in the wave function appear at the points of greatest positive and negative gradients of the potential. This phenomenon might be utilized in atom interferometry as a beam-splitter for matter waves\cite{mazzoni2015large}.

\begin{figure}
\begin{center}
\includegraphics[width=0.5\textwidth]{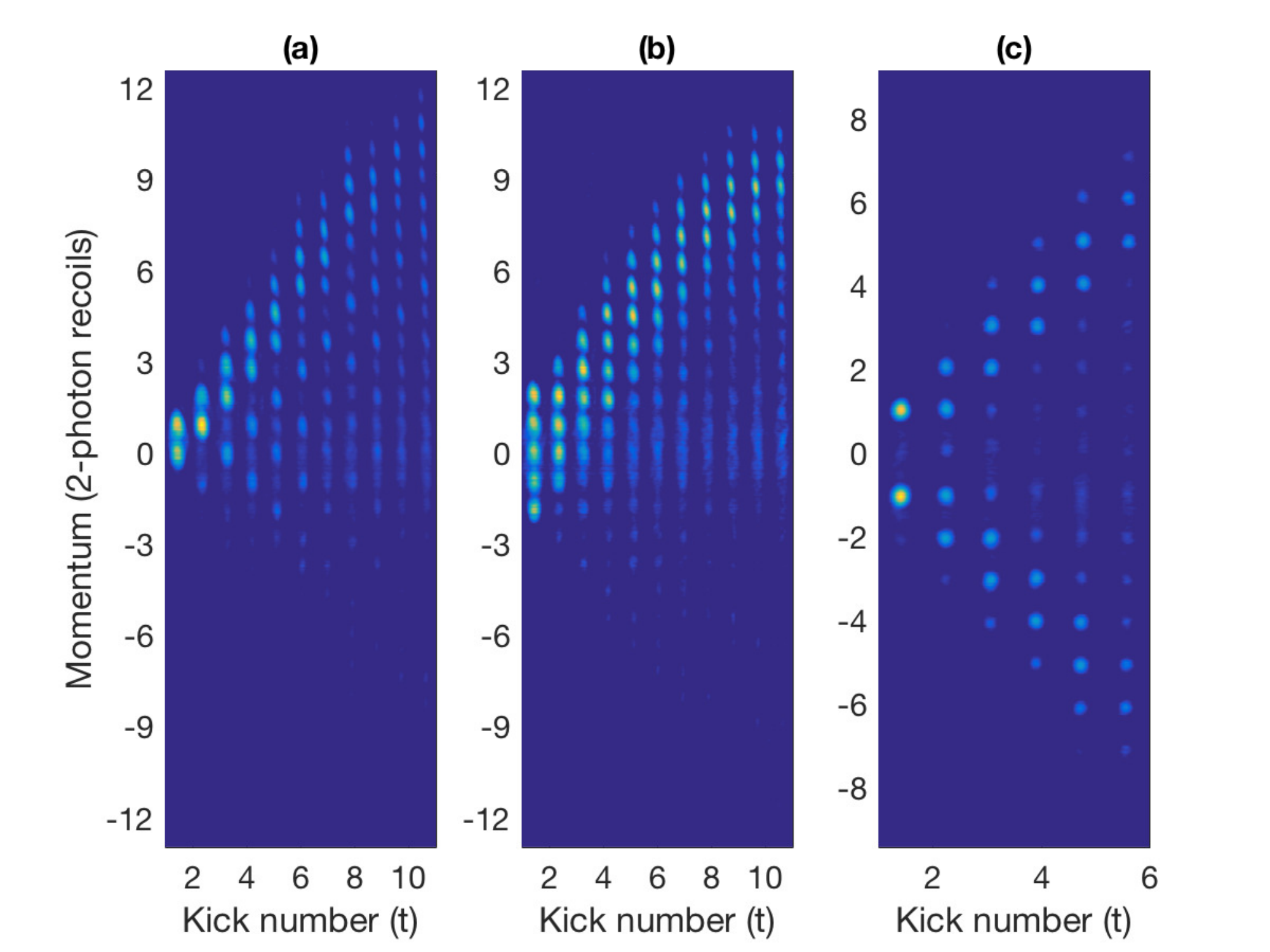} 
\caption{Experimental time-of-flight images for ratchets starting from initial states with different combinations of momentum states. The momentum current evolution of the initial combinations: $|0\rangle+|1\rangle$, $\sum_{n=-2}^{2}|n\rangle$, and $|-1\rangle+|1\rangle$ are shown versus kick number $t$ in panels (a)-(c) respectively. The relative phases of all momentum components in the initial states were set to zero but effectively implemented by offsetting the phase of the standing wave accordingly, see main text.}
\label{ }
\end{center}
\end{figure}

\subsection{Data analysis}
\label{sec:2c}

To understand the behavior of ratchets with different initial states a comprehensive study of the ratchet ``dispersion'' was experimentally conducted as a function of several variables. Here we define the normalized dispersion as:
\begin{equation}
\label{ }
D(t)=\frac{\langle p_{t}^{2}\rangle-\langle p_{t}\rangle^{2}}{\langle p_{0}^{2}\rangle-\langle p_{0}\rangle^{2}},
\end{equation}
in which, $t$ signifies the kick number ($t=0$ corresponds to the initial distribution), $\langle p_{t}\rangle$ is the mean momentum, $\langle p_{t}^{2}\rangle$ is the mean momentum squared at time $t$. The dispersion analysis is an objective approach to describe the ``quality'' of the ratchet current; the closer $D(t)$ is to $1$, the more the ratchet state resembles the initial superposition. Reducing the amount by which the dispersion increases can be essential for experiments such as the realization of discrete-time quantum walks \cite{GW2016}. The sensitivity of the ratchet current dispersion to the initial state was examined, the results of which are shown in Figure 6(a). \red{The experimental results are in strong agreement with what one would infer from FWHM of the wave functions shown in Fig. 2(a). That is, a narrower FWHM produces a much slower increase in dispersion with kick number. As one might expect, a larger number of momentum components in the initial state results in a better ratchet.}
\begin{figure}
\begin{center}
\includegraphics[width=0.5\textwidth]{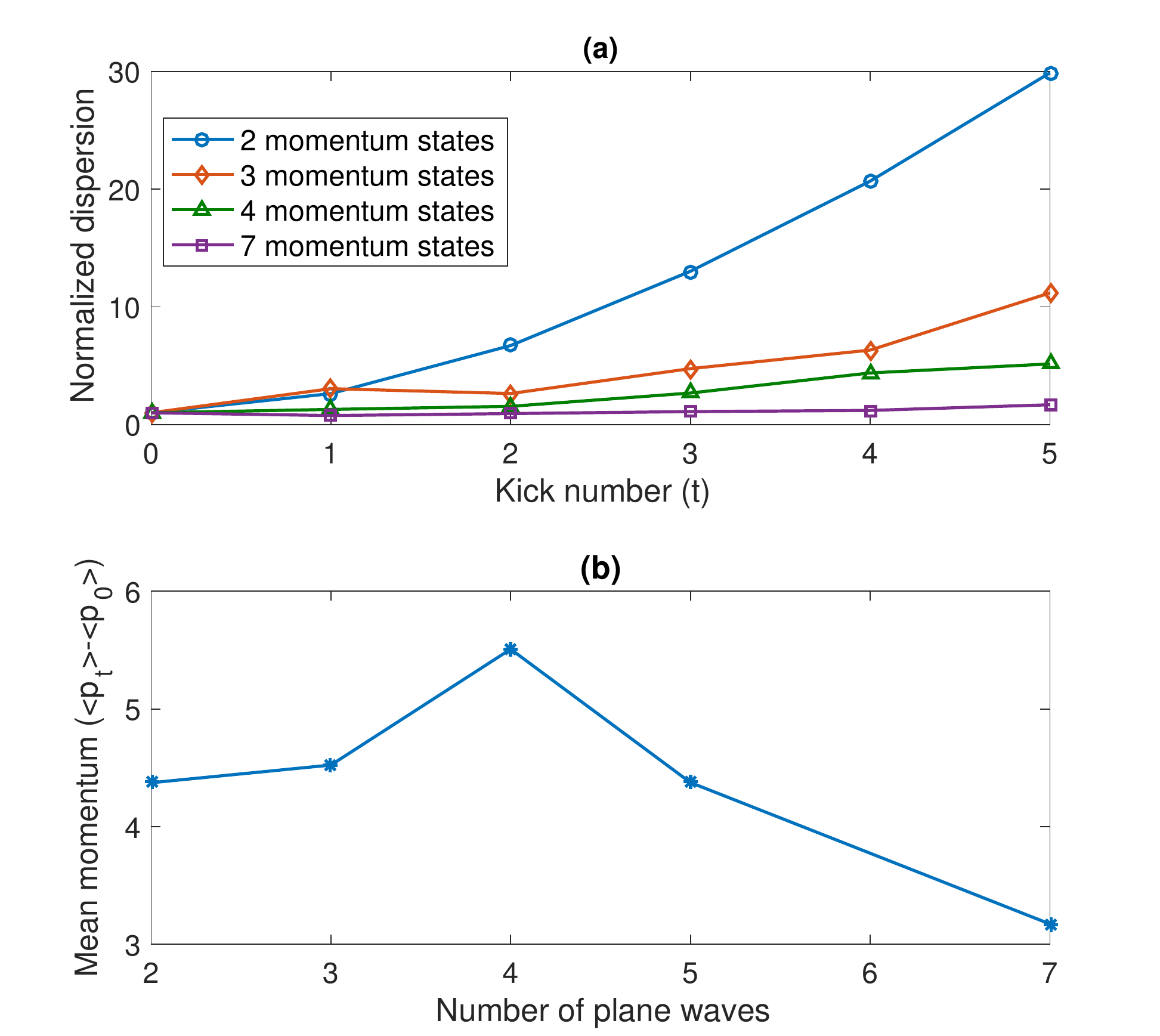}
\caption{Panel (a) illustrates the experimental results of the variation of the normalized dispersion versus number of kick with $\beta=0.5$ and $T=T_{1/2}$ for ratchets with initial states $\sum_{n=-3}^{3}|n\rangle$ (squares), $\sum_{n=-1}^{2}|n\rangle$ (triangles), $\sum_{n=-1}^{1}|n\rangle$ (diamonds), and $\sum_{n=0}^{1}|n\rangle$ (circles). Panel (b) illustrates the experimental results of the variation of the mean momentum versus the number of plane waves with $t=5$, $\beta=0.5$, and $T=T_{1/2}$ for ratchets. Momenta are plotted in units of two-photon recoils. Error bars are small \red{compared to the plotted point size} and are not shown here. \red{Panel (a) from \cite{ni2016}.}}
\label{fig:6}
\end{center}
\end{figure}

Mean momentum is another approach to examine the quality of the ratchet current; the larger the mean momentum, the stronger the ratchet current will be \cite{dana2008experimental}. The sensitivity of the ratchet mean momentum to the initial state was therefore investigated. The results shown in Figure 7 plot the mean momentum with kick number for $\beta=0.5$ and $T=T_{1/2}$. The experimental results reveal that the mean momentum of the ratchets starting with a smaller number of momentum states undergoes an early saturation, leading to a reduced slope in the early part of their evolution. The difference of the mean momentum between initial states of 5 momentum states and 3 momentum states is not as obvious as the one between initial states of 3 momentum states and 2 momentum states. That is reasonable and can be explained by Fig. 2(b), which shows that the difference of the effective force between the initial states diminishes as the number of component states increases. Fig. 6(a) reveals the same type of behavior. Here the mean momentum $\langle p_{t}\rangle$ is taken relative to the initial mean momentum $\langle p_{0}\rangle$ which is not necessarily zero for all combinations of initial states (for example, the mean momentum of the initial state $|0\rangle +|1\rangle$ is $0.5$).
\begin{figure}
\begin{center}
\includegraphics[width=0.5\textwidth]{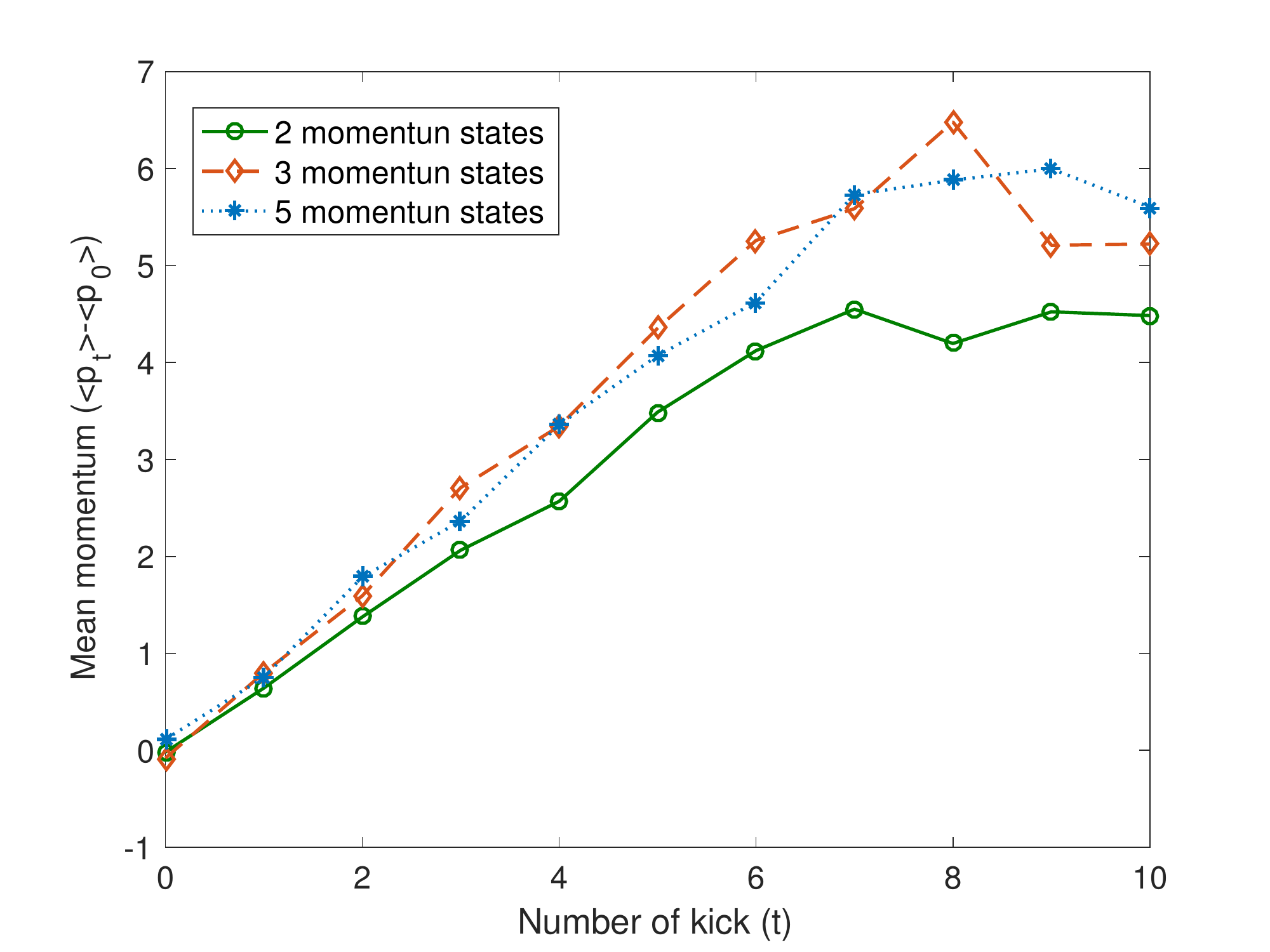}
\caption{Experimental results showing variation of the mean momentum versus kick number with $\beta=0.5$ and $T=T_{1/2}$ for ratchets with initial states $\sum_{n=-2}^{2}|n\rangle$ (asterisks), $\sum_{n=-1}^{1}|n\rangle$ (diamonds), and $\sum_{n=0}^{1}|n\rangle$ (circles). Momenta are plotted in units of two-photon recoils. Error bars are small \red{compared to the plotted point size} and not shown.}
\label{fig:7}
\end{center}
\end{figure}

As can be inferred from Figure 2(c), the ratchet behavior is dependent on the relative phase between the initial state and the $\delta$-kick standing wave potential. In this regard, sensitivity of the ratchet to the relative phase was examined at $t = 5$ for different initial states under similar kicking conditions ($\phi_{d}=1.4$, $\beta=0.5$, and $T=T_{1/2}$). Fig. 8 shows the corresponding experimental results for the ratchet dispersion and mean momentum as functions of the relative phase with initial states $\sum_{n=0}^{1}|n\rangle$ and $\sum_{n=-2}^{2}|n\rangle$. As can be seen in Fig. 8(a), the extrema of the dispersions occur at similar phases for both cases, whereas the amplitude for the initial state containing more momentum states is much smaller and shows less sensitivity to the phase parameter. The dependence of the mean momentum on the relative phase was also investigated for the two ratchets with these initial states. Corresponding experimental results are shown in Fig 8(b). As can be seen, while the extrema of the mean momenta occur for both ratchets at similar phases, the amplitude of mean momentum variation in either direction is considerably greater for the initial state containing more momentum states. In addition, it can be inferred from Fig. 8(a) and 8(b) that regardless of the number of initial state constituents, minimum dispersion and maximum mean momentum, ratchets occur at phases of $\pi/2$ and $3\pi/2$. However, compared to the mean momentum, the dispersion for different initial states is much more sensitive to the initial state.
\begin{figure}
\begin{center}
\includegraphics[width=0.5\textwidth]{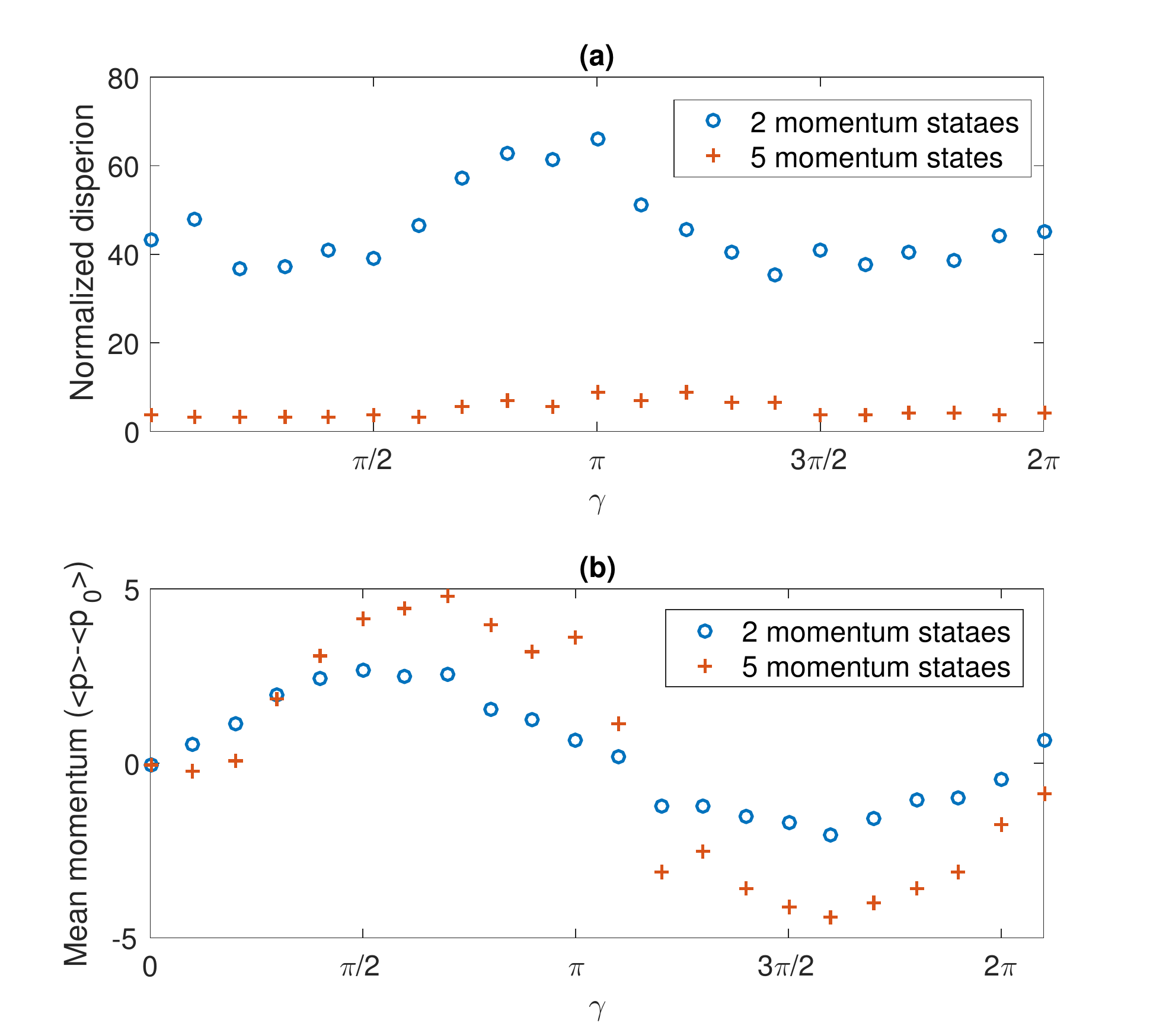}
\caption{Experimental results for the variation of (a) the normalized dispersion and (b) the mean momentum versus potential phase $\gamma$. For both panels, $t=5$, $\phi_{d}=1.4$, $\beta=0.5$, and $T=T_{1/2}$ for initial states $\sum_{n=-2}^{2}|n\rangle$ (pluses) and $\sum_{n=0}^{1}|n\rangle$ (circles). Momenta are plotted in units of two-photon recoils.}
\label{fig:8}
\end{center}
\end{figure}

\section{Off-resonance ratchet}
\label{sec:3}

Up to this point we have only considered ratchets formed by pulses separated by a quantum resonance time. However from an experimental perspective there is no need to be so restrictive. Theoretically studying the off-resonance ratchets with the initial state wave function approach used in the resonant case is very complicated. This is because the relative phase between the initial state wave function in position space and the $\delta$-kicks potential is not easily determined. In other words, the force exerted by the standing wave potential gradient on the atomic wave function, as the ratchet evolves can not be predicted from a simple model. Therefore investigating the influence of different variables on the evolution of an off-resonance ratchet is challenging.

The question of what happens to a ratchets away from resonance has been investigated in several theoretical studies \cite{Gong2008,sadgrove2009pseudo}. In ref.\cite{sadgrove2009pseudo} a pseudo-\-classical ratchet theory was developed and the existence of a one-\-parameter scaling law that could be used for prediction of the ratchet current for a wide variety of parameter was proposed. With this scaling law, the behavior of the ratchet could be described and controlled using a single universal variable containing many of the experimental parameters. In this section, the experimental realization of the off-resonance ratchet is discussed. By using a scaled mean momentum a single variable that depends on the kicking strength, time offset from resonance, kick number, and relative phase can describe the ratchet. It is also shown that momentum current inversions are possible for some ranges of this variable. For the sake of simplicity, the experiments were conducted only for the initial state consisting two momentum states, $|0\rangle+e^{-i\gamma}|1\rangle$.

\subsection{Theory}
To investigate off-resonant quantum ratchets, we start by writing the scaled pulse period as $\tau=2\pi l+\varepsilon$, where $|\varepsilon|$ measures the proximity to the resonance, and can be shown to play the role of Planck's constant \cite{fishman2002stable,fishman2003theory,wimberger2003quantum}. For this reason, at small values of $|\varepsilon|$ the dynamics of the off-resonance quantum ratchet can be understood by the classical mapping \cite{fishman2002stable,sadgrove2005ballistic},
\begin{equation}
\left\{
\begin{array}{rcl}
  J_{t+1} &=& J_{t}+\tilde{k}\sin(\theta_{t+1}) \\
  \theta_{t+1} &=& \theta_{t}+J_{t}.
\end{array}\right.
\end{equation}
Here $\tilde{k}=|\varepsilon|\phi_{d}$ signifies the scaled kicking strength, $J_{t}=\varepsilon p_{t}+l\pi+\tau\beta$ is the scaled momentum variable, and $\theta_{t}=(Gx)$mod$(2\pi)+\pi[1-sign(\varepsilon)]/2$ is the scaled position exploiting the spatial periodicity of the kick potential. As mentioned above, the off-resonance ratchet experiments were carried out using an initial superposition of two plane waves $|\psi\rangle=|0\rangle+e^{-i\gamma}|1\rangle$, that is equivalent to a rotor state $1+e^{i(\theta-\gamma)}$. This leads to the population distribution $\phi(\theta)=|\psi|^{2}=1+\cos(\theta-\gamma)$ in angular position space. As in quantum resonance, $\gamma$ is an additional phase used to account for the distance by which the initial spatial atomic distribution can be shifted relative to the standing wave potential. Although the distribution $\phi(\theta)$ is quantum in origin, it will be interpreted from now on as a classical distribution. 

The original application of $\varepsilon$-classical theory to the kicked rotor system showed the existence of a one-parameter scaling law for the mean energy of a kicked rotor \cite{wimberger2003quantum}. This prediction was experimentally verified in the vicinity of the first and second quantum resonances ($l=1$ and $l=2$) in Ref. \cite{wimberger2005experimental}. The existence of a one-parameter scaling law was also proposed for the ratchet current \cite{sadgrove2009pseudo}. One of the notable features of this theory is that it predicts an inversion of momentum current at some values of the scaling variable, that is, for certain families of real parameters.

The motion of the kicked rotor in continuous time can be described in the pendulum approximation \cite{casati1984non} by the scaled Hamiltonian $H'\approx(J')^{2}/2+|\varepsilon|\phi_{d}\cos(\theta)$, where $J'=J/(\sqrt{\phi_{d}|\varepsilon|})$ is a scaled momentum. In the vicinity of quantum resonance, one can calculate $\langle J'-J'_{0}\rangle=\int^{\pi}_{\pi}d\theta_{0}|\phi(\theta_{0})|^{2}(J'-J'_{0})$ with $\phi(\theta)$ the wave function in angular position space. Eq. (7) gives a phase space map dominated by a pendulum-like resonance island of extension $4\sqrt{\tilde{k}}\gg|\varepsilon|$ \cite{wimberger2003quantum}. Hence $p=0$ and $p=1$ essentially contribute in the same way, giving $J'_{0}=0$ so that the map in Eq. (7) is $J'_{t+1}=\sqrt{\tilde{k}}\sum^{t=N-1}_{t=0}\sin(\theta_{t+1})$. As a result, the off-resonance ratchet starting with a superposition of two plane waves: $|\psi\rangle=|0\rangle+e^{-i\gamma}|1\rangle$ leads to the scaled momentum $\langle J'-J'_{0}\rangle=-\sin\gamma S(z)$, where
\begin{equation}\label{}
  S(z)=\frac{1}{2\pi}\int_{-\pi}^{\pi}\sin\theta_{0}J'(\theta_{0},J'_{0}=0,z)d\theta_{0}
\end{equation}
with $z=t\sqrt{\phi_{d}|\varepsilon|}$ is a scaling variable. Thus the mean momentum (in units of $\hbar G$) can be expressed in terms of the scaled quantities as
\begin{equation}\label{}
  \langle p\rangle=\sqrt{\frac{\phi_{d}}{|\varepsilon|}}\langle J'-J'_{0}\rangle=-\frac{\phi_{d}t\sin\gamma}{z}S(z).
\end{equation}
From which,
\begin{equation}\label{}
\frac{\langle p\rangle}{-\phi_{d}t\sin\gamma}=\frac{S(z)}{z},
\end{equation}
where $S(z)$ is computed from Eq. (8) \cite{sadgrove2009pseudo}.

\subsection{Experiments and results}
Experimental momentum distributions of the off-\-resonant ratchet are shown in Fig. 9(a) and 9(b) as a function of the pulse period's offset from the first quantum resonance, and the kick number, respectively. It can be seen that there are certain values of these parameters where the distributions are weighted towards negative momenta. This is strong evidence for the existence of the current reversals predicted by the theory \cite{sadgrove2009pseudo}. 
\begin{figure}
\begin{center}
\includegraphics[width=0.5\textwidth]{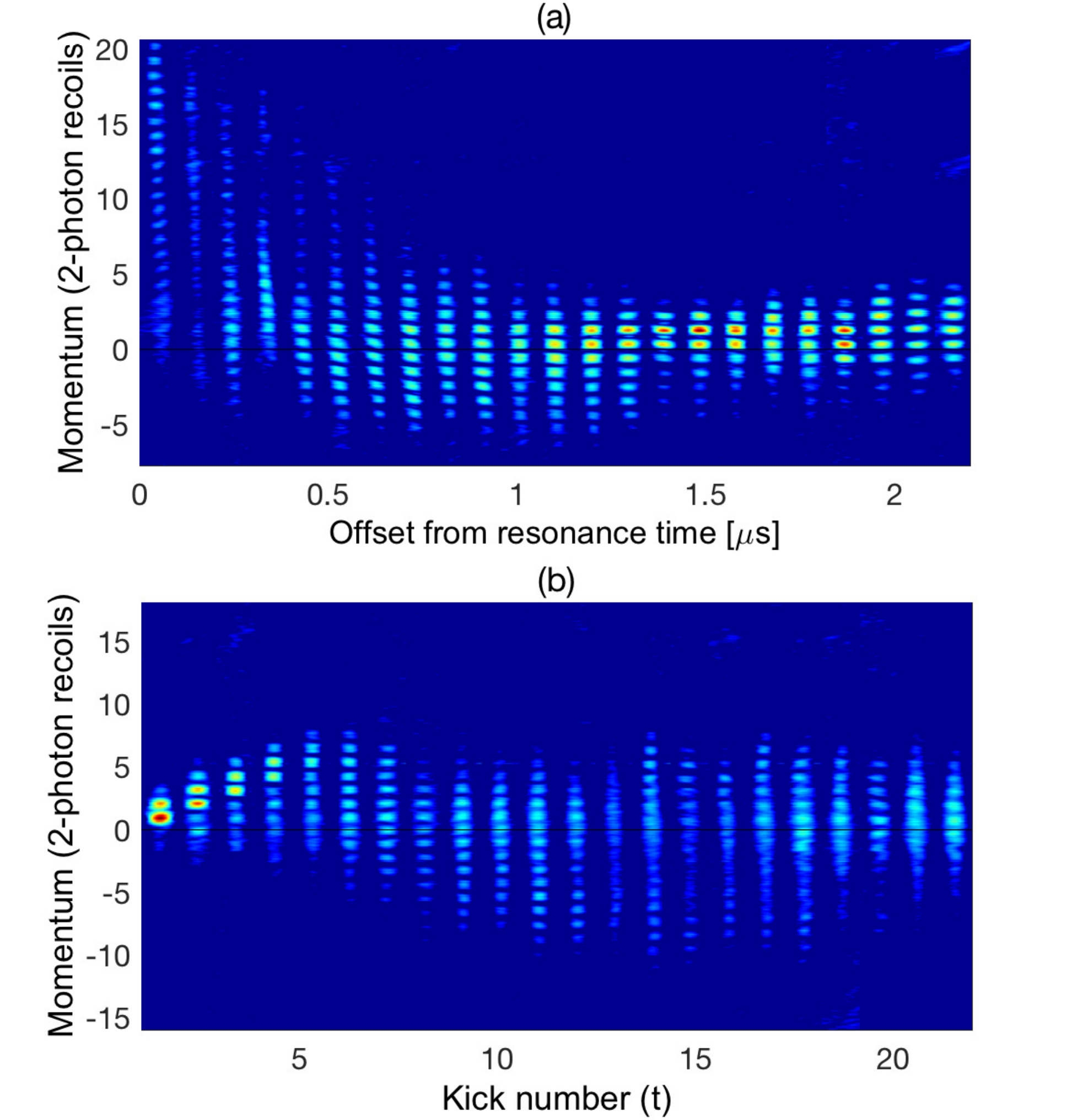}
\caption{Experimental momentum distributions after kicking the initial state $|0\rangle+e^{-i\gamma}|1\rangle$ by the short pulses of an off-resonant optical standing wave. The distributions are displayed as a function of (a) offset from resonance pulse period (10 kicks, $\phi_{d}=2.6$, and $\gamma=-\pi/2$) and (b) kick number ($|\varepsilon|=0.18$, $\phi_{d}=1.8$, and $\gamma=-\pi/2$). Image of each column was captured in a separate time-of-flight experiment. In both panels, weighting of the distributions are towards positive momenta at small values of each control variable followed by a tendency towards negative mean momenta around the middle of the horizontal axis range. In (b) the mean momenta tends asymptotically to 0 due to decoherence effects from spontaneous emission after about 15 kicks. \red{From \cite{shrestha2012controlling}.}}
\label{fig:9}
\end{center}
\end{figure}

The setup and procedure for the off-resonance  experiments were quite similar to those for quantum resonance, although the kicking pulse length was a little longer at 1.54 $\mu$s \cite{shrestha2012controlling,shrestha2013cold}. The experiments were carried out using values of $\tau$ close to the zeroth ($l=0$) and first ($l=1$) quantum resonances. Since $\tau$ plays the role of an effective Planck's constant, $\tau\rightarrow 0$ is the true classical limit \cite{sadgrove2005ballistic}. The measurements involved the determination of the mean momentum of the ratchet for various combinations of the parameters $t$, $\phi_{d}$, and $\varepsilon$. Figures 10 and 11 illustrate the variation of the measured momenta, scaled by $\phi_{d}t\sin\gamma$, as a function of the scaling variable $z$ for $l=0$ and $l=1$, respectively. In Fig. 10, $z$ was changed by varying the kick number $t$ for different combinations of $|\varepsilon|$ and $\phi_{d}$. In Fig. 11, $z$ was changed by scanning over each of $|\varepsilon|$, $\phi_{d}$, or $t$, while holding the other two constituents of $z$ constant. The solid line in both figures is the theoretical plot of the function $S(z)/z$ given by Eq. (10). It can be seen that, regardless of how $z$ is varied, the experimental results are in good agreement with the theory for various combinations of variables. There exists a regime over $z$ both cases where an inversion of the ratchet current takes place, with a maximum inversion at $z\approx5.6$. Interestingly this reversal of the ratchet takes place without shifting any of the centers of symmetry of the system. In addition, despite the fact that $\varepsilon$-classical theory is presumed to work only for small values of $|\varepsilon|$, the experimental results show that it can still be used successfully for higher values of this parameter. In fact, as expected from a Heisenberg-Fourier argument \cite{sadgrove2005ballistic,wimberger2005experimental,sadgrovea2011experiment}, the interval of valid $|\varepsilon|$ depends on the kick number $t$ \cite{wimberger2003quantum}, being relatively large for small $t\leq10$-$15$.
\begin{figure}
\begin{center}
\includegraphics[width=0.5\textwidth]{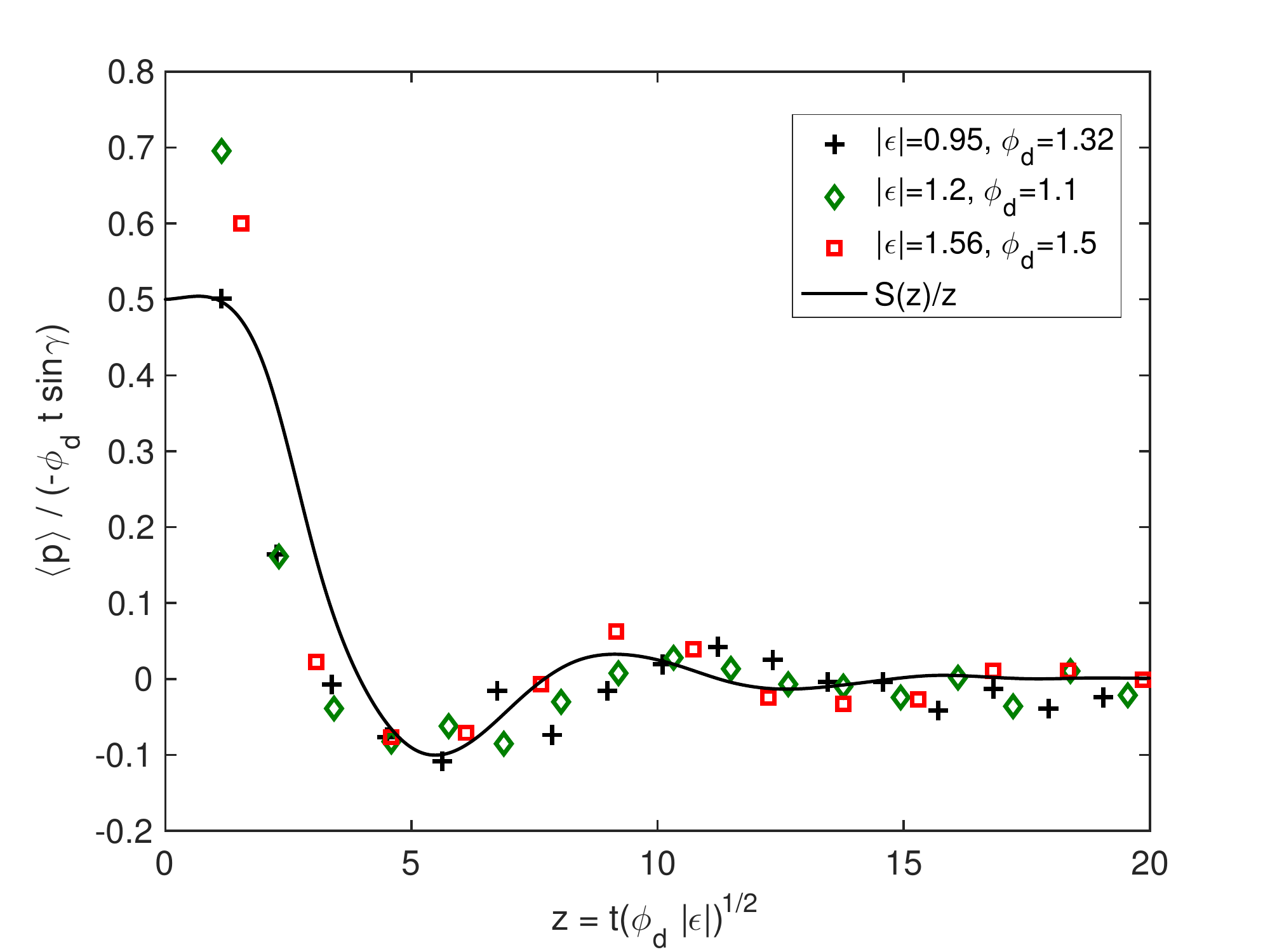}
\caption{Scaled mean momentum $\langle p\rangle/(\phi_{d}t\sin\gamma)$ as a function of the dimensionless scaling variable $z=\sqrt{\phi_{d}|\varepsilon|}t$ for $l=0$. $z$ was varied by scanning over kick number $t$ for different combinations of $\phi_{d}$ and $|\varepsilon|$. The solid line is the function $S(z)/z$ given by Eq. (10). \red{Adapted from \cite{shrestha2012controlling}.}}
\label{fig:10}
\end{center}
\end{figure}

\begin{figure}
\begin{center}
\includegraphics[width=0.5\textwidth]{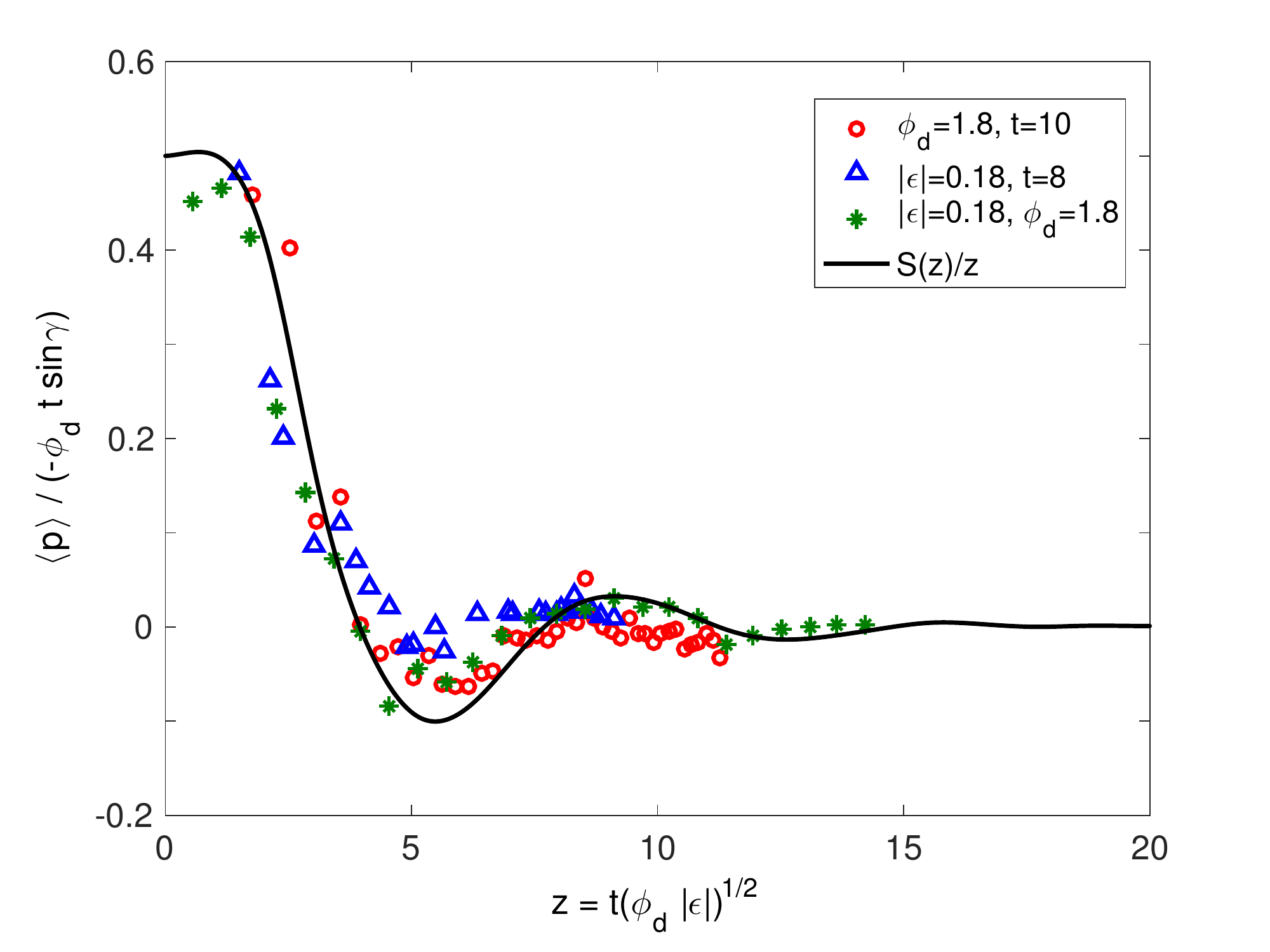}
\caption{Scaled mean momentum $\langle p\rangle/(\phi_{d}t\sin\gamma)$ as a function of the scaling variable $z=\sqrt{\phi_{d}|\varepsilon|}t$ for $l=1$. $z$ was varied by scanning over $|\varepsilon|$ with $t=10$ and $\phi_{d}=1.8$ (circles), and $\phi_{d}$ with $|\varepsilon|=0.18$ and $t=8$ (triangles), and $t$ with $|\varepsilon|=0.18$ and $\phi_{d}=1.8$ (asterisks). The solid line is the function $S(z)/z$ given by Eq. (10). This demonstrates that no matter how $z$ is obtained the scaled mean momentum is approximately universal. \red{Adapted from \cite{shrestha2012controlling}.}}
\label{fig:11}
\end{center}
\end{figure}

\section{Conclusion}
\label{sec:4}

In this paper we have reviewed our work on the experimental implementation of quantum-resonance ratchets based on periodically kicked BECs with attention given to the quality of the ratchet currents. In particular, we studied the dispersion, see  Fig. \ref{fig:6}, as well as the maximum net directed current produced for various initial states and relative phases, see Figs. \ref{fig:7} and \ref{fig:8}, respectively. Also the behavior of the ratchet was investigated for a detuning in the drive from the precise resonance conditions. Directed currents of matter-waves could be useful for future applications in interferometry with larger momentum differences \cite{mazzoni2015large, Kasevich} as well as in the implementation of discrete-time quantum walks over a substantial number of steps \cite{GW2015, GW2016}. Indeed we are already working on the realization of such optimized quantum walks controlled by the entanglement between internal atomic degrees of freedom and their center-of-mass momenta \cite{GW2016}.


\end{document}